\documentclass[english,aps,preprint]{revtex4}
\usepackage[T1]{fontenc}
\usepackage[latin9]{inputenc}
\usepackage{color}
\usepackage{float}
\usepackage{amsmath}
\usepackage{graphicx}
\usepackage{amssymb}
\usepackage{esint}

\makeatletter
\def\romannumber#1{\@roman{#1}}
\def\Romannumber#1{\@Roman{#1}}

\@ifundefined{showcaptionsetup}{}{%
 \PassOptionsToPackage{caption=false}{subfig}}
\usepackage{subfig}
\makeatother

\usepackage{babel}

\begin{document}
\thispagestyle{empty}

\begin{center}
\textbf{Diffusion of cold atomic gases}
\par\end{center}

\smallskip{}

\begin{center}
\textbf{in the presence of an optical speckle potential}
\par\end{center}

\bigskip{}

\begin{center}
\textbf{L. Beilin, E. Gurevich and B. Shapiro }
\par\end{center}

\bigskip{}

\begin{center}
Department of Physics, Technion - Israel Institute of Technology,
Haifa 32000, Israel\\

\par\end{center}

\pagestyle{plain} 
\begin{abstract}
We consider diffusion of a cold Fermi gas in the presence of a random
optical speckle potential. The evolution of the initial atomic cloud
in space and time is discussed. Analytical and numerical results are
presented in various regimes. Diffusion of a Bose-Einstein condensate
is also briefly discussed and similarity with the Fermi gas case is
pointed out.
\end{abstract}
\maketitle

\section{{\normalsize Introduction }}

Transport of cold atomic gases in the presence of a quenched random
potential is a rapidly developing field of research \cite{Fallani}.
In a typical set-up the gas is released from a harmonic trap and undergoes
expansion, while being scattered by the random potential. At some
later time an image of the expanded atomic cloud is taken and, thus,
information about the mode of transport (ballistic, diffusive or localized)
is obtained. The random potential for atoms is obtained by creating
a random pattern of light intensity (optical speckle). Experiments
on propagation of cold atoms through optical speckles have been limited
so far to one-dimensional ($1d$) geometry and have culminated in
observation of $1d$ Anderson localization for a Bose-Einstein condensate
(BEC) \cite{Billy,Roati}.

There is a considerable amount of theoretical work on diffusion and
possible localization of an expanding BEC cloud in two and three dimensions
\cite{Shapiro-1,Miniatura-1,Skipetrov-1,Cherroret,Schwiete,Cherroret-Skipetrov}.
The same problem can be also addressed for a cold Fermi gas - a system
which is intensively studied in recent years (see \cite{Giorgini-1,Castin}
for recent reviews). Diffusion of an expanding Fermi gas, in the long
time limit and for a Gaussian white noise potential, was discussed
in \cite{Hensler-1}. In the present paper we consider the experimentally
relevant case of a speckle potential, concentrating on $2d$ geometry.
In Sec. $\Romannumber{2}$ we write down the basic equations which
govern the evolution of a diffusing Fermi cloud. In Sec. $\Romannumber{3}$
we summarize, following \cite{Kuhn}, the behavior of the diffusion
coefficient $D(k)$, as a function of the particle wave number $k$,
in a $2d$ speckle potential. In Sec. $\Romannumber{4}$ we study
the density $n(\vec{r},t)$ of a diffusing Fermi gas as a function
of position and time. Since $n(\vec{r},t)$ is expressed by an integral
which cannot be calculated analytically, we resort to numerics in
combination with an analytic treatment of some limiting cases. In
Sec. $\Romannumber{5}$ we briefly discuss the evolution of the shape
of a diffusing BEC and point out some similarities (and differences)
with the case of the Fermi gas.

\section{Basic Equations}

We consider $N$ fermions at zero temperature, initially trapped in
a harmonic potential. At time $t=0$ the trap is switched off, while
a random potential $V(\vec{r})$ is switched on. Our aim is to study
the dynamics of the atoms, upon their release from the trap, in the
presence of the random potential. In many circumstances interactions
between the fermions have only a minor effect on their dynamics. This
is particularly true for a polarized Fermi gas when the Pauli principle
eliminates the main mechanism (the s-scattering) for the interaction.
In the absence of interactions the single particle wave functions,
$\Psi_{n}(\vec{r},t),$ describing individual atoms, evolve according
to: \begin{equation}
i\hbar\partial_{t}\Psi_{n}(\vec{r},t)=-\frac{\hbar^{2}}{2m}\triangle\Psi_{n}(\vec{r},t)+V(\vec{r})\Psi_{n}(\vec{r},t)\:,\label{eq:Schredinger equation}\end{equation}
with the initial condition $\Psi_{n}(\vec{r},0)=\Phi_{n}(\vec{r})$,
where $\Phi_{n}(\vec{r})$ is $n$'th eigenstate of the harmonic potential
$\frac{1}{2}m\omega^{2}r^{2}$, and $V(\vec{r})$ is the random potential,
with zero mean and a two-point correlation function $\left\langle V\left(\vec{r}_{1}\right)V\left(\vec{r}_{2}\right)\right\rangle =\Gamma\left(\vec{r}_{2}-\vec{r}_{1}\right)$.
\\
The formal solution of (\ref{eq:Schredinger equation}) is\begin{equation}
\Psi_{n}(\vec{r},t)=\int d\vec{R}G(\vec{r},\vec{R},t)\Phi_{n}(\vec{R})\:,\end{equation}
where $G(\vec{r},\vec{R},t)$  is the retarded Green\textquoteright{}s
function of the Schrödinger equation (\ref{eq:Schredinger equation}).
The quantum expectation value of the particle density (per one spin
component) at time $t$ and for a given realization of randomness
is\begin{equation}
<\hat{n}(\vec{r},t)>=\sum_{n}f_{n}\left|\Psi_{n}(\vec{r},t)\right|^{2}=\int d\vec{R}d\vec{R}'G^{*}(\vec{r},\vec{R},t)G(\vec{r},\vec{R'},t)\sum_{n}f_{n}\Phi_{n}^{*}(\vec{R})\Phi_{n}(\vec{R}')\:,\label{eq:diffusion expectation value}\end{equation}
where $f_{n}$ is the occupation function, which for zero temperature
is given by the step function $\Theta(E_{F}-E_{n})$. Averaging $<n(\vec{r},t)>$
over the disorder yields \begin{equation}
<\overline{\hat{n}(\vec{r},t)}>=\int d\vec{R}d\vec{R}'\overline{G^{*}(\vec{r},\vec{R},t)G(\vec{r},\vec{R'},t)}\sum_{n}f_{n}\Phi_{n}^{*}(\vec{R})\Phi_{n}(\vec{R}')\:.\label{eq:diffusion expectation value averaged1}\end{equation}
In order to average the product of the two Green's functions in (\ref{eq:diffusion expectation value averaged1})
we first Fourier transform to the energy representation\begin{equation}
\overline{G^{*}(\vec{r},\vec{R},t)G(\vec{r},\vec{R}',t)}=\int\frac{d\varepsilon}{2\pi}\int\frac{d\Omega}{2\pi}e^{-\frac{i\Omega t}{\hbar}}\overline{G^{*}(\vec{r},\vec{R},\varepsilon+\frac{1}{2}\Omega)G(\vec{r},\vec{R'},\varepsilon-\frac{1}{2}\Omega)}\:.\label{eq: product of 2 G}\end{equation}
The product in r.h.s. of (\ref{eq: product of 2 G}), in the diffusion
approximation, is represented diagrammatically in Fig.1. %
\begin{figure}[H]
\centering{}\includegraphics[scale=0.75]{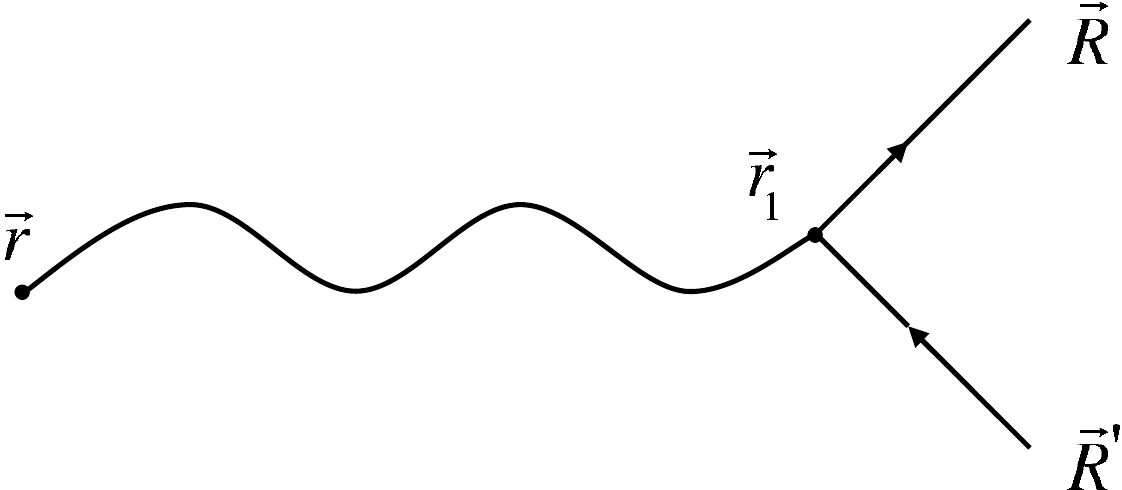}\caption{Diagrammatic representation of the product in Eq.(\ref{eq: product of 2 G}).}

\end{figure}
 The two straight lines represent a wave and its complex conjugate
propagating from their sources ($\vec{R'}$ and $\vec{R}$) to an
intermediate point $\vec{r}_{1}$. At this point the two waves {}``recombine''
and the wave intensity propagates by diffusion to the observation
point $\vec{r}$. The corresponding algebraic expression is \begin{equation}
\overline{G^{*}(\vec{r},\vec{R},\varepsilon+\frac{1}{2}\Omega)G(\vec{r},\vec{R'},\varepsilon-\frac{1}{2}\Omega)}=\frac{\hbar}{\tau_{\varepsilon}}\int d\vec{r}_{1}P_{\varepsilon}(\vec{r},\vec{r_{1}},\Omega)\overline{G^{*}}(\vec{r}_{1}-\vec{R},\varepsilon+\frac{1}{2}\Omega)\overline{G}(\vec{r}_{1}-\vec{R'},\varepsilon-\frac{1}{2}\Omega)\:,\label{eq:product of 2 G 1}\end{equation}
where $\tau_{\varepsilon}$ is the scattering mean free time at energy
$\varepsilon$, $P_{\varepsilon}(\vec{r},\vec{r}_{1},\Omega)$ is
the diffusion ladder\cite{Akkermans} and $\overline{G}(\vec{r},\varepsilon)$
is the average Green's function. For the latter $\Omega$ can be neglected,
in comparison with $\varepsilon$, and its explicit expression is
\begin{equation}
\overline{G}(\vec{r},\varepsilon\pm\frac{1}{2}\Omega)\approx\overline{G}(\vec{r},\varepsilon)=G_{0}(\vec{r},\varepsilon)e^{-\frac{r}{2l_{\varepsilon}}}\quad,\label{eq:the average Green function}\end{equation}
where $G_{0}$ is the free Green\textquoteright{}s function and $l_{\varepsilon}$
is the single particle mean free path. Since the Green's functions
in Eq.(\ref{eq:product of 2 G 1}) rapidly decay (at a distance $l_{\varepsilon}$),
the slow varying diffusion ladder $P_{\varepsilon}(\vec{r},\vec{r}_{1},\Omega)$
can be taken out of the integral, with the argument $\vec{r}_{1}$
being replaced by $\frac{\vec{R}+\vec{R}'}{2}$. Performing the remaining
integral and returning to (\ref{eq: product of 2 G}) yields \begin{equation}
\overline{G^{*}(\vec{r},\vec{R},t)G(\vec{r},\vec{R}',t)}=-\frac{1}{\pi}\int d\varepsilon P_{\varepsilon}\left(\vec{r},\frac{\vec{R}+\vec{R}'}{2},t\right)Im\overline{G}\left(\vec{R}-\vec{R}',\varepsilon\right)\:,\label{eq:Product of 2 G 2}\end{equation}
where the diffusion propagator \begin{equation}
P_{\varepsilon}(\vec{r},\vec{R},t)=\frac{1}{\left(4\pi D_{\varepsilon}t\right)^{d/2}}\exp\left(-\frac{\left|\vec{r}-\vec{R}\right|^{2}}{4D_{\varepsilon}t}\right)\label{eq:diffusion prop}\end{equation}
is the Fourier transform of $P_{\varepsilon}(\vec{r},\vec{R},\Omega)$
and $D_{\varepsilon}$ is the diffusion coefficient at energy $\varepsilon$.
The necessary condition for the above derivation is $kl_{\varepsilon}\gg1$,
where $k=\sqrt{2m\varepsilon/\hbar^{2}}$. \\
Substituting (\ref{eq:Product of 2 G 2}) into (\ref{eq:diffusion expectation value averaged1})
and using the fact that for weak disorder $-\frac{1}{\pi}Im\overline{G}(\vec{k},\varepsilon)\simeq\delta(\varepsilon-\varepsilon_{k})$
one obtains\textit{\textcolor{red}{ }}\\
\begin{equation}
<\overline{\hat{n}(\vec{r},t)}>\equiv n(\vec{r},t)=\int d\vec{R}\int d\vec{p}P_{p}(\vec{r},\vec{R},t)\sum_{n}f_{n}W_{n}(\vec{p},\vec{R})\:,\label{eq:diffusion expectation value averaged2}\end{equation}
where $P_{p}$ is given by (\ref{eq:diffusion prop}) with $\varepsilon=\frac{p^{2}}{2m}$
and\begin{equation}
W_{n}(\vec{p},\vec{R})\equiv\frac{1}{(2\pi\hbar)^{d}}\int d\vec{\rho}e^{\frac{i}{\hbar}\vec{p}\vec{\rho}}\Phi_{n}^{*}(\vec{R}+\frac{1}{2}\vec{\rho})\Phi_{n}(\vec{R}-\frac{1}{2}\vec{\rho})\:\end{equation}
 is the Wigner transform of $\Phi_{n}(\vec{r})$ . In the classical
limit \textcolor{black}{($n\gg1$)} the Wigner function $W_{n}(\vec{p},\vec{R})$
for an eigenstate $n$ becomes \cite{Quantum Optics} \begin{equation}
W_{n}(\vec{p},\vec{R})=\frac{1}{(2\pi\hbar)^{d}\nu(E_{n})}\delta(E_{n}-\frac{p^{2}}{2m}-\frac{1}{2}m\omega^{2}R^{2})\:,\end{equation}
where $E_{n}$ is the energy of state $n$ and $\nu(E_{n})$ is the
density of states for a particle in a harmonic trap. Substituting
this into (\ref{eq:diffusion expectation value averaged2}) and replacing
summation over $n$ by integration over energy up to the Fermi energy
$E_{F}$ we finally obtain \begin{equation}
n(\vec{r},t)=\int d\vec{R}\int\frac{d\vec{p}}{(2\pi\hbar)^{d}}P_{p}(\vec{r},\vec{R},t)\Theta(E_{F}-\frac{p^{2}}{2m}-\frac{1}{2}m\omega^{2}R^{2}).\label{eq:Density Fermi 2}\end{equation}

\section{{\normalsize Diffusion coefficient in speckle disorder}}

In order to proceed with the evaluation of the integral in (\ref{eq:Density Fermi 2}),
an explicit expression is required for the diffusion coefficient\begin{equation}
D(k)=\frac{\hbar kl_{B}}{dm}\:,\label{eq:Boltzmann diffusion coefficient D(k) in any dimension d}\end{equation}
where $l_{B}$ is the Boltzmann transport mean free path and $k=\frac{p}{\hbar}$.
So far we have not specified the type of disorder. Now we specialize
to a two-dimensional ($d=2$) speckle potential, generated by transmitting
laser light through circular diffusive plate, whose two-point correlation
function is given by \cite{Miniatura-1}\begin{equation}
\Gamma(\vec{r}_{1}-\vec{r}_{2})=4V_{0}^{2}\left[\frac{J_{1}(k_{0}\left|\vec{r}_{1}-\vec{r}_{2}\right|)}{k_{0}\left|\vec{r}_{1}-\vec{r}_{2}\right|}\right]^{2}\:,\label{eq:correlation function 2D}\end{equation}
where $J_{1}$ is the first-order Bessel function, $V_{0}$ is the
standard deviation and $k_{0}$ is the inverse correlation length
of the random potential. The latter is related to the laser wavelength
and numerical aperture of the imaging device. Then, in the weak disorder
limit, the mean free path is given by \cite{Kuhn}\begin{equation}
\frac{1}{kl_{B}}=\eta^{2}\left(\frac{k_{0}}{k}\right)^{2}\intop_{0}^{2\pi}\frac{1}{2\pi}d\theta\widetilde{\Gamma}\left(2\frac{k}{k_{0}}\left|\sin\left(\frac{\theta}{2}\right)\right|\right)(1-\cos(\theta))\:,\label{eq:Boltzman transport mean-free path 1}\end{equation}
where $\eta=\frac{V_{0}}{E_{0}}$ is the measure of the potential
fluctuations strength, $E_{0}=\frac{\hbar^{2}k_{0}^{2}}{m}$ is the
{}``correlation'' energy and\begin{equation}
\widetilde{\Gamma}\left(\kappa\right)=8\left(\arccos\left(\frac{\kappa}{2}\right)-\frac{\kappa}{2}\sqrt{\left(1-\left(\frac{\kappa}{2}\right)^{2}\right)}\right)\Theta\left(2-\kappa\right)\:.\end{equation}
In the limiting cases, $k\ll k_{0}$ and $k\gg k_{0}$, (\ref{eq:Boltzman transport mean-free path 1})
may be approximated as \cite{Kuhn}: \begin{equation}
kl_{B}\approx\left\{ \begin{array}{c}
\frac{1}{4\pi\eta^{2}}\left(\frac{k}{k_{0}}\right)^{2}\;\;\;\;,\;\;\;\; k\ll k_{0}\\
\frac{45\pi}{128\eta^{2}}\left(\frac{k}{k_{0}}\right)^{5}\;\;\;,\;\;\;\; k\gg k_{0}\end{array}\right.\:.\label{eq:Boltzmann transport mean free path 2}\end{equation}
In Fig. 2(a) we compare approximations (\ref{eq:Boltzmann transport mean free path 2})
to the exact numerical evaluation of (\ref{eq:Boltzman transport mean-free path 1}).
The optimal choice of a point, separating between the two asymptotics,
is the crossing point $k_{cr}=\lambda k_{0}$, with $\lambda=\left(\frac{32}{45\pi^{2}}\right)^{1/3}\approx0.41$.
With this choice (\ref{eq:Boltzman transport mean-free path 1}) is
approximated as:\begin{equation}
kl_{B}=\left\{ \begin{array}{c}
\frac{1}{4\pi\eta^{2}}\left(\frac{k}{k_{0}}\right)^{2}\;\;\;\;,\;\;\;\; k<\lambda k_{0}\\
\frac{45\pi}{128\eta^{2}}\left(\frac{k}{k_{0}}\right)^{5}\;\;\;,\;\;\;\; k>\lambda k_{0}\end{array}\right.\:.\label{eq:Boltzman transport mean free path 3}\end{equation}
The approximation (\ref{eq:Boltzman transport mean free path 3})
differs from the exact numerical solution of (\ref{eq:Boltzman transport mean-free path 1})
by a numerical factor of order unity. This is demonstrated in Fig.
2(b), which shows the ratio between the two. %
\begin{figure}[H]
\subfloat[]{\includegraphics[scale=0.5]{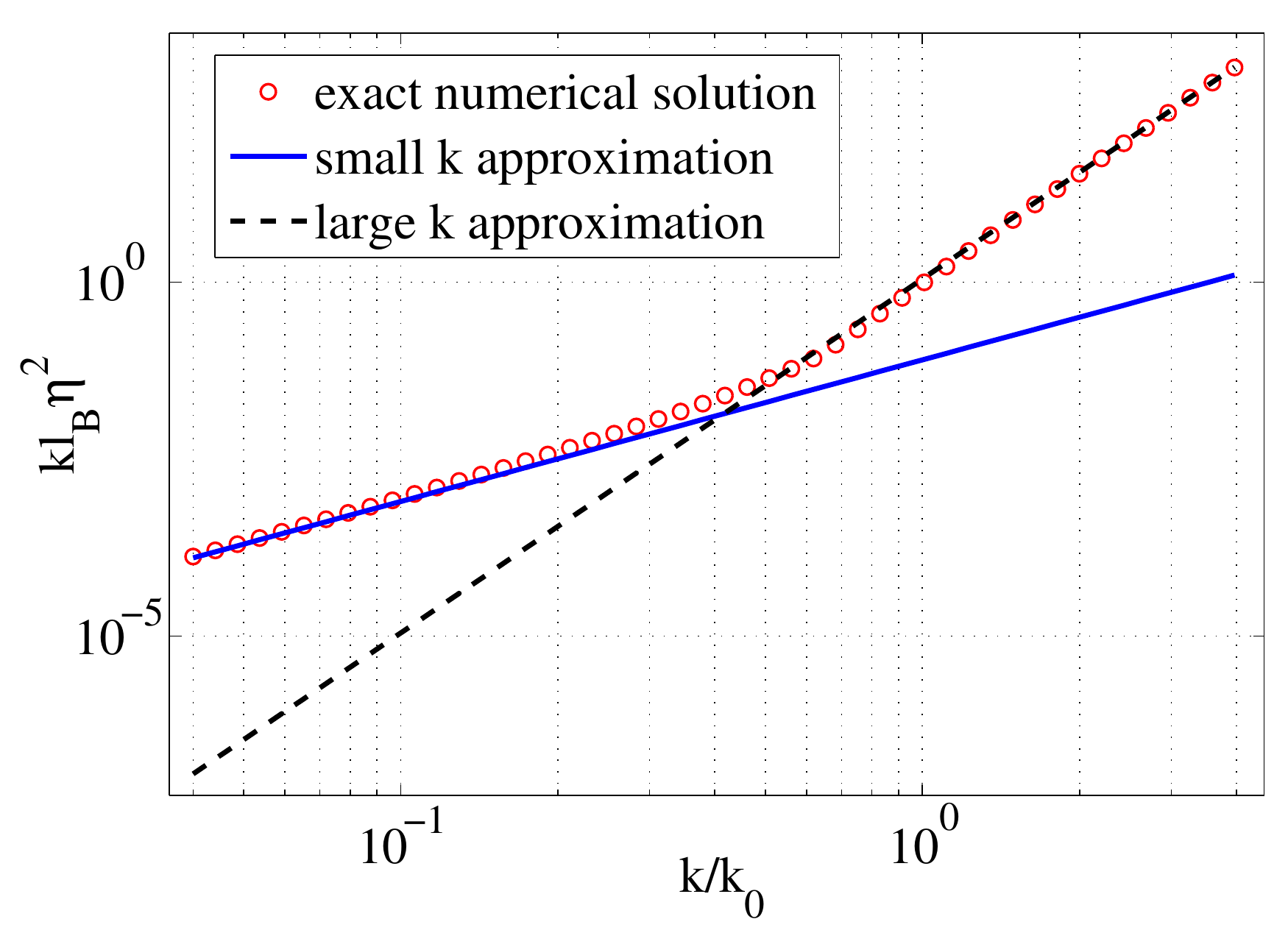}

}\subfloat[]{\includegraphics[scale=0.5]{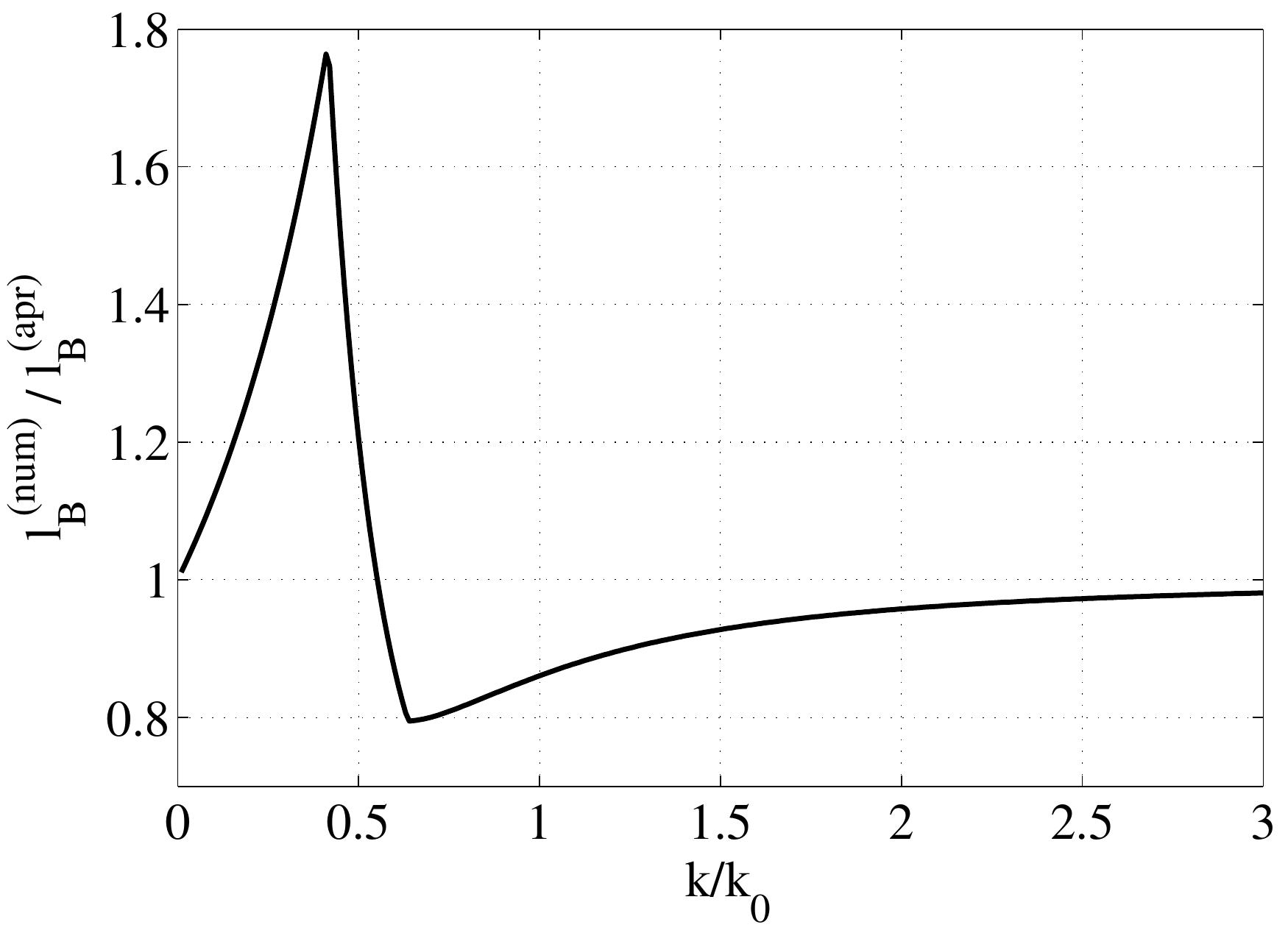}

}\caption{(Color online) (a) Comparison between the numerical (\textcolor{red}{$\circ$}),
Eq.\ref{eq:Boltzman transport mean-free path 1}, and approximate
(lines), Eq.\ref{eq:Boltzman transport mean free path 3}, solution
for the Boltzmann transport mean free path $l_{B}$. The small $k$
(solid line) and large $k$ (dashed line) asymptotics cross at $k_{cr}=\lambda k_{0}$;
(b) The ratio between the exact numerical solution of (\ref{eq:Boltzman transport mean-free path 1})
and the approximate expression (\ref{eq:Boltzman transport mean free path 3}).
The maximal deviation, obtained at the crossing point $k_{cr}=\lambda k_{0}$,
is about $1.76$.}

\end{figure}
Thus, we write the diffusion coefficient as \begin{equation}
D(k)=\left\{ \begin{array}{c}
D_{<}(k)=\frac{D_{0}}{8\pi}\left(\frac{k}{k_{0}}\right)^{2}\;\;\;\;\;,\;\;\; k_{c}<k<\lambda k_{0}\\
D_{>}(k)=\frac{45\pi D_{0}}{256}\left(\frac{k}{k_{0}}\right)^{5}\;\;,\;\;\;\;\;\;\;\;\; k>\lambda k_{0}\end{array}\right.,\label{eq:Boltzman 2d final}\end{equation}
where $D_{0}\equiv\frac{\hbar^{5}k_{0}^{4}}{m^{3}V_{0}^{2}}$. Note
that we have introduced a lower cutoff $k_{c}$, which is derived
from the Ioffe-Regel criterion $k_{c}l_{B}=1$ (see below). For $k<k_{c}$
the diffusion approximation employed in this paper is not valid any
more: thus, particles with $k<k_{c}$ remain localized in the vicinity
of the initial trap.

\section{{\normalsize Evolution of the density in space and time}}

\textcolor{black}{Using the above explicit expression for the diffusion
coefficient, one can calculate the atomic density profile $n(\vec{r},t)$
(\ref{eq:Density Fermi 2}).} It is convenient to introduce the following
dimensionless variables:\begin{equation}
\left\{ \begin{array}{c}
\widetilde{r}=\frac{r}{R_{Max}}\\
\widetilde{t}=\frac{t}{t_{0}}\\
\widetilde{n}=nR_{Max}^{2}\end{array}\right.\:,\label{eq:Rescaled units in 2D}\end{equation}
where $R_{Max}=\sqrt{\frac{2E_{F}}{m\omega^{2}}}=\frac{\sqrt{8N}}{k_{F}}\;$\textcolor{black}{
is the initial size of the atomic cloud and $t_{0}=\frac{R_{Max}^{2}}{D_{0}}=\frac{2\eta^{2}\sqrt{2N}}{\omega}$
is a characteristic diffusion time.} Let us note that for $t\rightarrow0$
the diffusion kernel (\ref{eq:diffusion prop}) becomes a delta function
$\delta(\vec{r}-\vec{R})$ and the density approaches its initial
shape of the inverted parabola, \begin{equation}
\widetilde{n}_{0}(\widetilde{r},0)=\frac{2N}{\pi}(1-\widetilde{r}^{2})\:,\end{equation}
\textcolor{black}{which corresponds to the Thomas-Fermi approximation
for $N$  fermions in the harmonic trap. Since it is difficult to
calculate analytically the integral in the expression (\ref{eq:Density Fermi 2}),
}below we consider various special cases.

In the long time limit the atomic cloud will spread to a distance
much larger than its initial size $R_{Max}$. Then, one can set $R=0$
in the diffusion kernel in (\ref{eq:Density Fermi 2}) and integrate
over $\vec{R}$, with the following result:\begin{equation}
n(\vec{r},t)=\int\frac{d\vec{p}}{(2\pi\hbar)^{2}}P_{p}(\vec{r},0,t)\left|\widetilde{\Phi}(p)\right|^{2}\,,\label{eq:Long time limit}\end{equation}
where \begin{equation}
\left|\widetilde{\Phi}(p)\right|^{2}=\pi R_{Max}^{2}\left(1-\frac{p^{2}}{p_{F}^{2}}\right)\Theta\left(p_{F}-p\right)\end{equation}
is the momentum distribution of the gas. Eq. (\ref{eq:Long time limit})
has a simple interpretation: it describes classical diffusion of particles
with momentum $\vec{p}$ and energy $\varepsilon=\frac{p^{2}}{2m}$
and with a momentum dependent diffusion coefficient given in (\ref{eq:Boltzman 2d final}).
It is interesting to note that (\ref{eq:Long time limit}) is completely
analogous to the corresponding expression for a diffusing BEC, with
$k_{F}=p_{F}/\hbar$ being replaced by the inverse healing length
$1/\xi$ \cite{Shapiro-1,Miniatura-1}.

The integral in (\ref{eq:Long time limit}) cannot be calculated analytically
due to the complicated dependence of the diffusion kernel on the particle
momentum $p$. A considerable simplification occurs if one assumes
$p_{F}\ll\hbar k_{0}.$ In this case all atomic wave numbers satisfy
the condition $k<k_{0}$ so that correlations in the random potential
do not come into play. The diffusion coefficient is given by $D_{<}(k)$
(see (\ref{eq:Boltzman 2d final})) in the whole range of integration
which corresponds to the limit of an uncorrelated, white-noise potential.
The expression (\ref{eq:Long time limit}) reduces to:\begin{equation}
n(r,t)=\frac{R_{Max}^{2}}{8\pi t}\int_{k_{c}}^{k_{F}}\frac{kdk}{D_{<}(k)}\exp\left(-\frac{r^{2}}{4D_{<}(k)t}\right)\left(1-\frac{k^{2}}{k_{F}^{2}}\right).\label{eq:White noise limit}\end{equation}
Let us stress that the white noise limit, Eq. (\ref{eq:White noise limit}),
requires that the typical strength $V_{0}$ of the random potential
must be smaller than the correlation energy $E_{0}$, so that the
parameter $\eta=\frac{V_{0}}{E_{0}}\ll1$ \cite{Miniatura-1}. Indeed,
the white noise condition, $k\ll k_{0},$ is compatible with the weak
disorder requirement, $kl_{B}>1$, only if $\eta\ll1$ (see (\ref{eq:Boltzman transport mean free path 3})).
This inequality implies $k_{c}=\sqrt{4\pi}\eta k_{0}\ll k_{0}$. Furthermore,
in order for the weak disorder requirement to be satisfied for the
great majority of the fermions, we must require $k_{F}\gg k_{c}$,
i.e. $E_{F}\gg\frac{V_{0}^{2}}{E_{0}}$. Switching to the dimensionless
variables and performing the integral yields: \begin{equation}
\widetilde{n}(\widetilde{r},\widetilde{t})=\frac{2Ns}{\widetilde{t}}\left[-e^{-\frac{\pi s\widetilde{r}^{2}}{\widetilde{t}}}+2\pi s\eta^{2}e^{-\frac{1}{\eta^{2}}\frac{\widetilde{r}^{2}}{2\widetilde{t}}}+\left(1+\frac{\pi s\widetilde{r}^{2}}{\widetilde{t}}\right)\left(E_{1}\left[\frac{\pi s\widetilde{r}^{2}}{\widetilde{t}}\right]-E_{1}\left[\frac{\widetilde{r}^{2}}{2\eta^{2}\widetilde{t}}\right]\right)\right]\:,\label{eq:E_0>2E_F one integral Final with mu}\end{equation}
where $s=\frac{E_{0}}{E_{F}}$ and the special function $E_{1}\left(x\right)$
is the exponential integral \cite{Math_Reference}. The aforementioned
condition $E_{F}\gg\frac{V_{0}^{2}}{E_{0}}$ implies that the parameter
$s\eta^{2}\ll1$. As an experimentally relevant example, we consider
the $Li^{6}$ atoms in the isotropic trap with the harmonic confinement
frequency $\frac{\omega}{2\pi}\approx160\, Hz$ and the speckle scale
$\frac{2\pi}{k_{0}}=0.5\mu m$. For $\eta=0.05$ and $s=12$, this
corresponds to $N\sim10^{4}$ atoms trapped in the initial cloud of
the radius $R_{max}\sim50\mu m$ and the typical time $t_{0}\sim0.7ms$,
which is about two orders of magnitude larger than the Boltzmann transport
mean free time $\tau_{B}$. Expression (\ref{eq:E_0>2E_F one integral Final with mu})
is plotted in Fig. 3 for $s$ and $\eta$ specified above. In Fig.
3(a) $\widetilde{n}(\widetilde{r},\widetilde{t})/N$ is shown as a
function of normalized time and distance. The chopped part of the
plot corresponds to the region where the approximation of long time
limit is not valid. Fig. 3(b) depicts snapshots of the density at
different times. {\large }%
\begin{figure}[H]
\subfloat[]{\includegraphics[bb=45bp 238bp 549bp 603bp,scale=0.5]{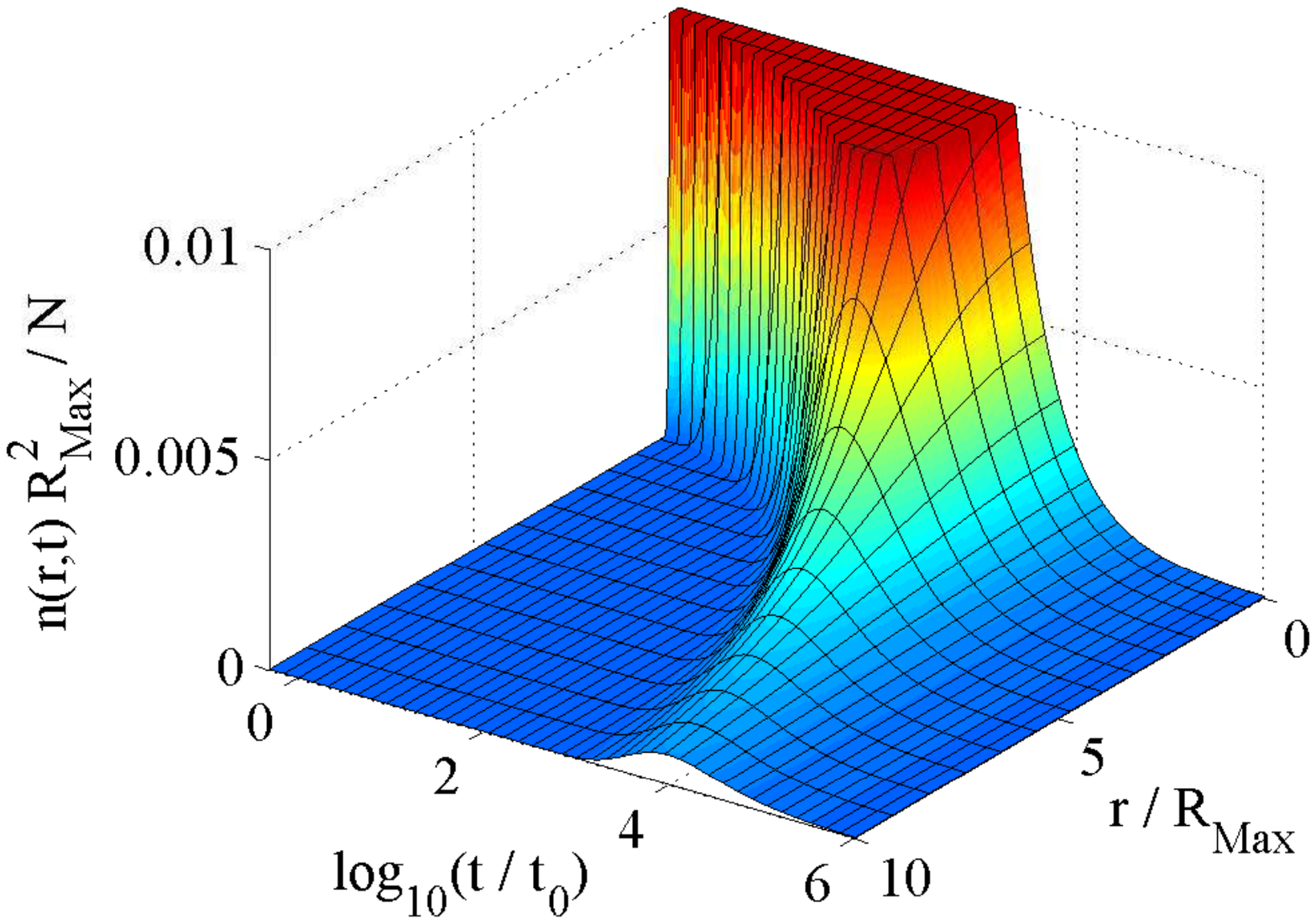}

}\subfloat[]{\includegraphics[bb=45bp 238bp 549bp 603bp,scale=0.5]{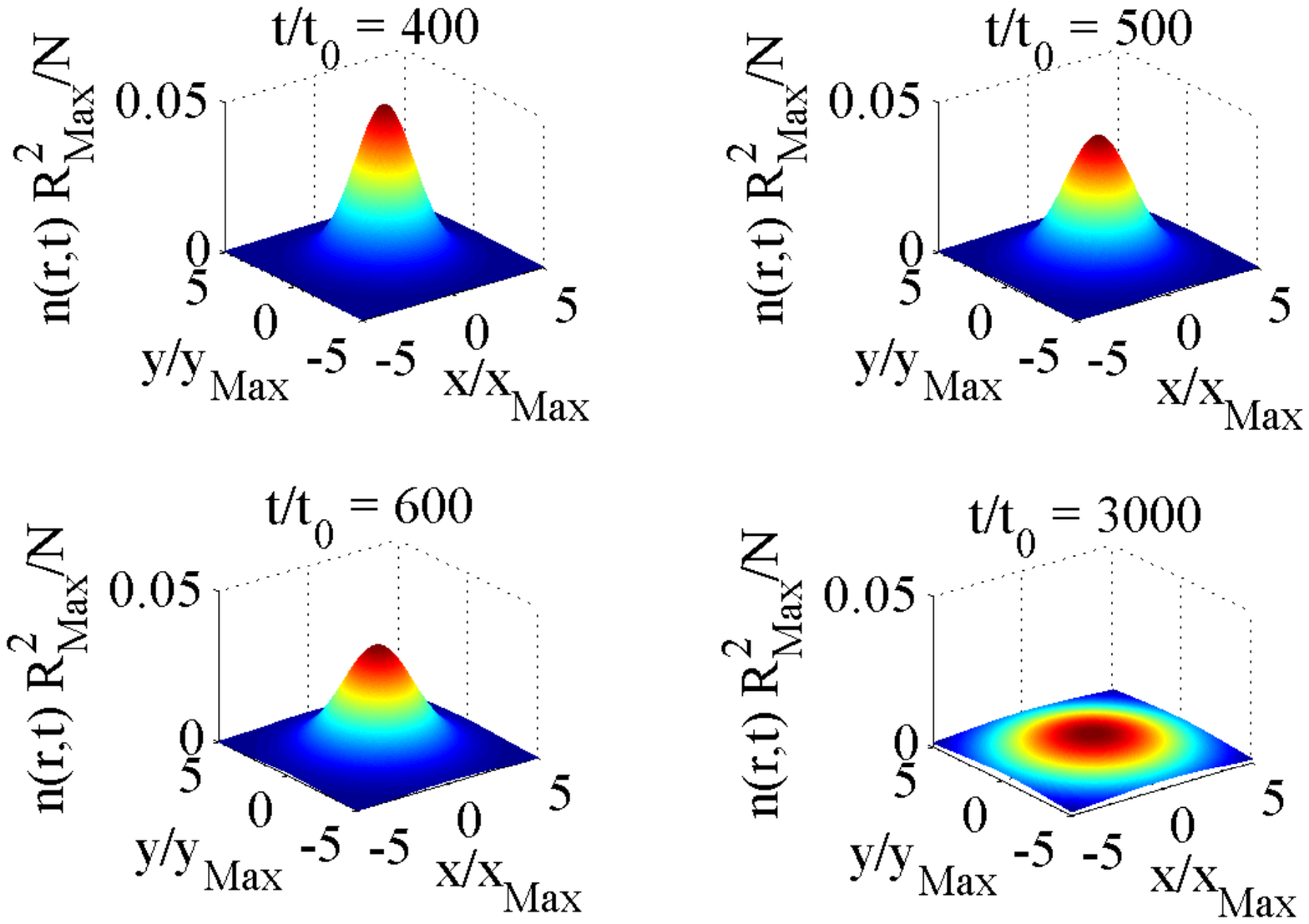}

}

{\large \caption{(Color online) (a) Normalized atomic density $\widetilde{n}(\widetilde{r},\widetilde{t})/N$
Eq.(\ref{eq:E_0>2E_F one integral Final with mu}), as a function
of normalized time and distance for $s=12$ and $\eta=0.05$. The
chopped part of the plot corresponds to small times and $\widetilde{r}$
for which the approximation of long time limit is not valid; (b) snapshots
of the density for $s=12$ and $\eta=0.05$ at four times $\frac{t}{t_{0}}=400,500,600,3000$.}
}
\end{figure}
Let us discuss the obtained expression (\ref{eq:E_0>2E_F one integral Final with mu})
in different regimes. For $r\lesssim R_{Max}$ (i.e. $\widetilde{r}\lesssim1$)
and for large times $\widetilde{t}>\frac{1}{\eta^{2}}$, using the
expansion of $E_{1}(x)$ for small values of $x$ \cite{Math_Reference}
\begin{equation}
E_{1}(x)=-\ln x+O(1)\:,\label{eq:Gamma Small Assimptotics}\end{equation}
Eq.(\ref{eq:E_0>2E_F one integral Final with mu}) simplifies to\begin{equation}
\widetilde{n}(\widetilde{r},\widetilde{t})\approx\frac{2Ns}{\widetilde{t}}\ln\left[\frac{1}{2\pi s\eta^{2}}\right]\:.\label{eq:t more 1/eta from approximation}\end{equation}
In the main region, $R_{Max}<r<\sqrt{4D_{<}(k_{F})t}$ (i.e. $1<\widetilde{r}<\sqrt{\frac{\widetilde{t}}{\pi s}}$),
expression (\ref{eq:E_0>2E_F one integral Final with mu}) is not
intuitive and, for the visualization, in Fig. 4 we compare it with
the solution for constant $D=D_{<}\left(k_{F}\right)$(see Eq.(\ref{eq:Boltzman 2d final}))\begin{equation}
\widetilde{n}(\widetilde{r},\widetilde{t})=\frac{Ns}{\widetilde{t}}e^{-\frac{\pi s\widetilde{r}^{2}}{\widetilde{t}}}(1-2\pi s\eta^{2})^{2},\label{eq:D constant 2D E0moreEF Final}\end{equation}
where the factor in the parentheses accounts for the lower momentum
cutoff $k_{c}.$ As expected, the solution for $2d$ speckle has a
more compact shape and the density decays faster than in the case
of constant $D$, for which the density shape is Gaussian.%
\begin{figure}[H]
\begin{centering}
\includegraphics[scale=0.5]{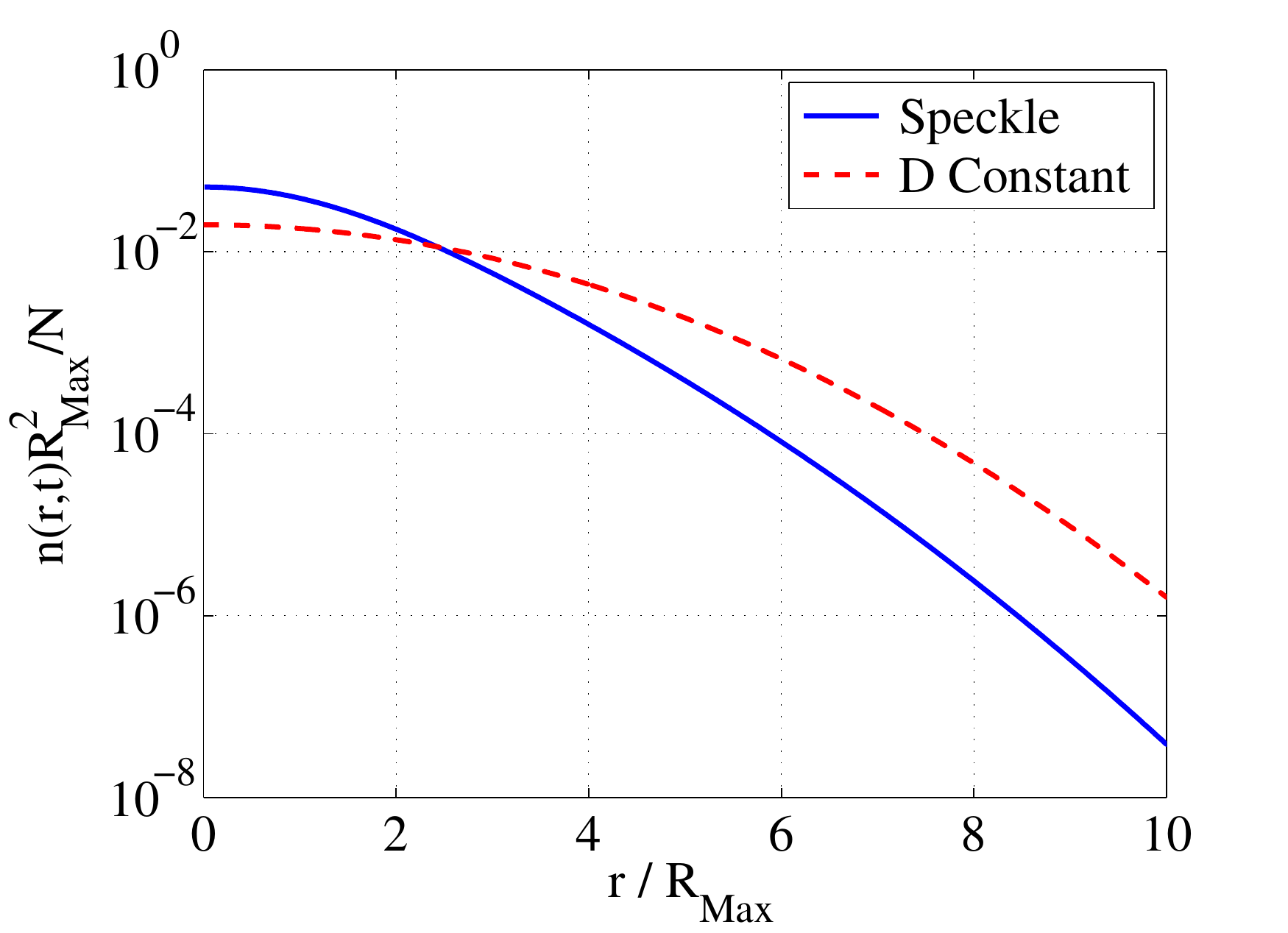}
\par\end{centering}

\caption{(Color online) Comparison between the Fermi gas density for $2d$
speckle, Eq. (\ref{eq:E_0>2E_F one integral Final with mu}), and
constant diffusion coefficient $D$, Eq. (\ref{eq:D constant 2D E0moreEF Final}),
at $\widetilde{t}=400$. The parameters are $s=12$ and $\eta=0.05$,
for which the momentum cutoff $p_{c}\ll p_{F}$. }

\end{figure}
For larger time, such that $r\ll\sqrt{4D_{<}(k_{F})t}$ (i.e. $\widetilde{r}\ll\sqrt{\frac{\widetilde{t}}{\pi s}}$),
and for $r>R_{Max}$, Eq.(\ref{eq:E_0>2E_F one integral Final with mu}),
with the help of (\ref{eq:Gamma Small Assimptotics}), reduces to
\begin{equation}
\widetilde{n}(\widetilde{r},\widetilde{t})\thickapprox\frac{2Ns}{\widetilde{t}}\ln\left[\min\left[\frac{\widetilde{t}}{\pi s\widetilde{r}^{2}},\frac{1}{2\pi s\eta^{2}}\right]\right]\:,\end{equation}
which differs from the {}``usual'' large time $1/t$ behavior by
the logarithmic factor. The later originates from the diffusion constant
dispersion. Finally, for $r>\sqrt{4D_{<}(k_{F})t}$ (and for $r<\frac{4D_{<}(k_{F})t}{R_{max}}$,
where (\ref{eq:E_0>2E_F one integral Final with mu}) is still valid),
one can use the large $x$ asymptotic expansion \cite{Math_Reference}\begin{equation}
E_{1}(x)=x^{-1}e^{-x}[1+O(\frac{1}{x})]\label{eq:Gamma Large assymptotics}\end{equation}
 to obtain \begin{equation}
\widetilde{n}(\widetilde{r},\widetilde{t})\approx\frac{2N\widetilde{t}}{\widetilde{r}^{4}s\pi^{2}}\exp\left(-\frac{\pi s\widetilde{r}^{2}}{\widetilde{t}}\right)\:,\end{equation}
which differs from the Gaussian decay by the algebraic factor $\frac{1}{\widetilde{r}^{4}}$.
\textcolor{black}{Let us note that this asymptotics is for zero temperature,
i.e. when there is a sharp cutoff of the atomic momentum distribution
at $E_{F}$. }

In order for the condition $k_{F}\ll k_{0}$ to be fulfilled, the
number of atoms $N$ should be fairly small. When $N$ increases,
for a fixed frequency trap $\omega$, one arrives to the opposite
regime $k_{F}\gg k_{0}.$ The integral (\ref{eq:Long time limit})
is then split into two parts (using (\ref{eq:Boltzman 2d final})):\begin{gather}
n(r,t)=\frac{R_{Max}^{2}}{8\pi t}\left[\int_{k_{c}}^{\lambda k_{0}}\frac{kdk}{D_{<}(k)}\exp\left(-\frac{r^{2}}{4D_{<}(k)t}\right)\left(1-\frac{k^{2}}{k_{F}^{2}}\right)\right.+\nonumber \\
+\left.\int_{\lambda k_{0}}^{k_{F}}\frac{kdk}{D_{>}(k)}\exp\left(-\frac{r^{2}}{4D_{>}(k)t}\right)\left(1-\frac{k^{2}}{k_{F}^{2}}\right)\right]\equiv n_{<}(r,t)+n_{>}(r,t)\:.\label{eq:Two integrals}\end{gather}
The first part, $n_{<}(r,t)$, describes contribution of {}``slow''
particles which diffuse with the coefficient $D_{<}(k)\sim k^{2}$,
as in a white noise potential. The second part, $n_{>}(r,t)$, corresponds
to {}``fast'' particles for which correlations in the random potential
lead to a sharp increase in the diffusion coefficient, $D_{>}(k)\sim k^{5}.$
For an arbitrary $r$, the solution of (\ref{eq:Two integrals}) is
given by :

\begin{gather}
\widetilde{n}(\widetilde{r},\widetilde{t})=\frac{2Ns}{\widetilde{t}}\left(F_{1}\left(\frac{2\pi\widetilde{r}^{2}}{\widetilde{t}}\right)+F_{2}\left(\frac{64}{45\pi}\frac{\widetilde{r}^{2}}{\widetilde{t}}\right)\right),\label{eq:E_0_less2Ef_Final with mu}\end{gather}
where\begin{align*}
F_{1}\left(x\right) & =2\pi s\eta^{2}e^{-\frac{x}{4\pi\eta^{2}}}-\frac{\lambda^{2}s}{2}e^{-\frac{x}{\lambda^{2}}}+\left(1+\frac{sx}{2}\right)\left(E_{1}\left[\frac{x}{\lambda^{2}}\right]-E_{1}\left[\frac{x}{4\pi\eta^{2}}\right]\right),\\
F_{2}\left(x\right) & =\frac{64}{225\pi^{2}}\left[\frac{s}{2}x^{-\frac{1}{5}}\left(\Gamma\left[\frac{1}{5},\frac{x}{\lambda^{5}}\right]-\Gamma\left[\frac{1}{5},\left(\frac{s}{2}\right)^{\frac{5}{2}}x\right]\right)+x^{-\frac{3}{5}}\left(\Gamma\left[\frac{3}{5},\left(\frac{s}{2}\right)^{\frac{5}{2}}x\right]-\Gamma\left[\frac{3}{5},\frac{x}{\lambda^{5}}\right]\right)\right]\end{align*}
and $\Gamma\left(\alpha,z\right)$ is the incomplete Gamma function.
In Fig. 5 the density $\widetilde{n}(\widetilde{r},\widetilde{t})/N$
is plotted for $s=\frac{1}{2}$ and $\eta=0.01$ as a function of
the normalized time and distance. One can observe that for fixed $r$
the time evolution of the density exhibits a slight kink. It is due
to the division of particles into two groups - {}``fast'' ($k>\lambda k_{0}$)
and {}``slow'' ($k<\lambda k_{0}$). It is not clear whether this
is a genuine physical effect or an artifact of the approximation (\ref{eq:Boltzman 2d final})
for $D(k)$. %
\begin{figure}[H]
\begin{centering}
\includegraphics[bb=33bp 172bp 576bp 619bp,scale=0.5]{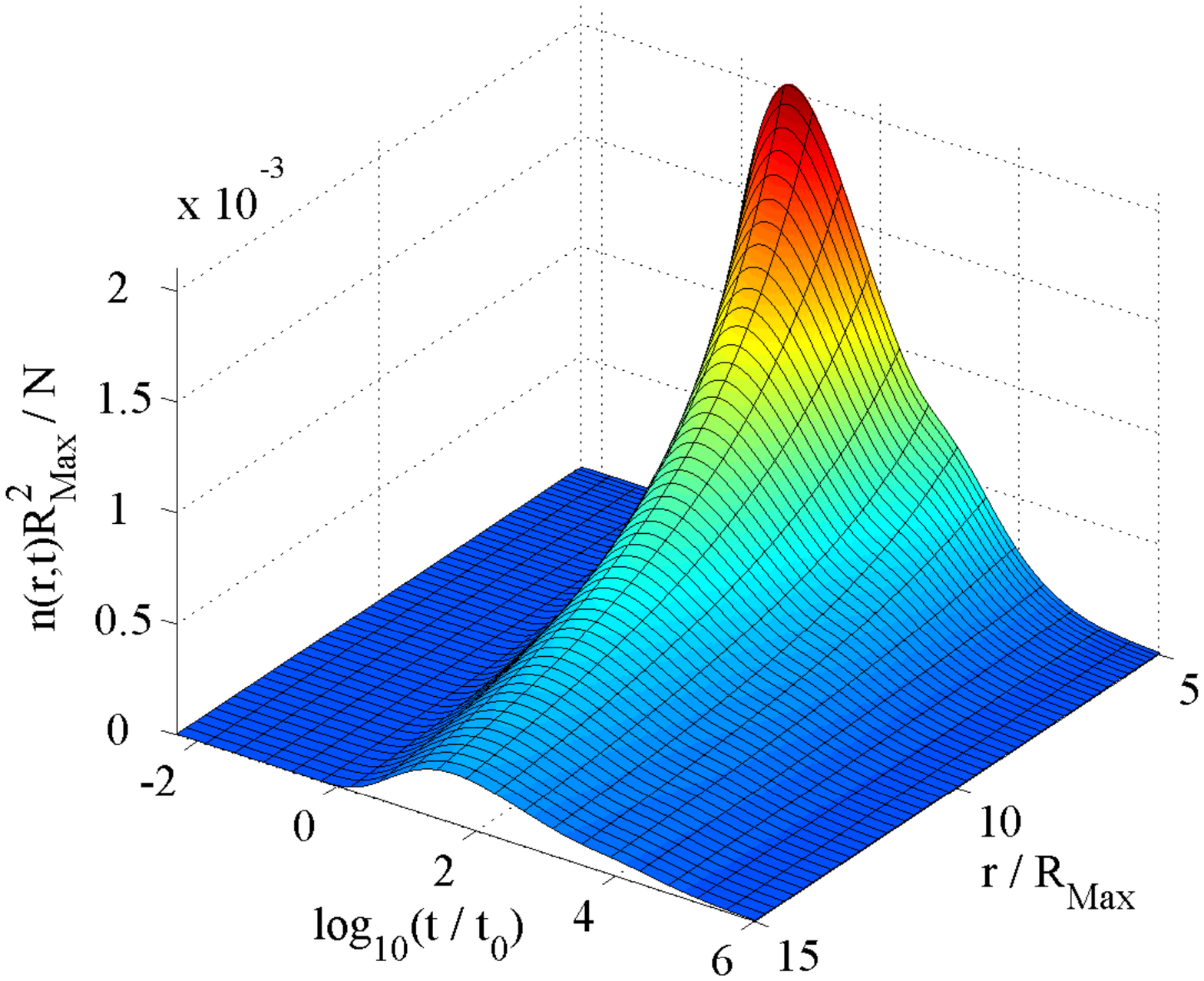}
\par\end{centering}

\caption{(Color online) Normalized atomic density $\widetilde{n}(\widetilde{r},\widetilde{t})/N$
Eq.(\ref{eq:E_0_less2Ef_Final with mu}) as a function of normalized
time and distance for $s=\frac{1}{2}$ and $\eta=0.01$. }

\end{figure}
For $r>\sqrt{4D_{>}(k_{F})t}$, the asymptotic tail of the solution
(\ref{eq:E_0_less2Ef_Final with mu}) can be written as \begin{equation}
\widetilde{n}(\widetilde{r},\widetilde{t})\sim\frac{1}{\widetilde{r}^{4}}\exp\left(-\beta s^{\frac{5}{2}}\frac{\widetilde{r}^{2}}{\widetilde{t}}\right)\,,\end{equation}
where $\beta=\frac{4}{45\pi\sqrt{2}}$.

The above discussion pertained to the case $\eta\ll1$. In the opposite
case, $\eta\gg1$, disorder correlations are important for all relevant
values of $k$, so that one should use $D_{>}(k)$ in the whole region
of integration. This is because the weak disorder condition, $kl_{B}>1$,
can now be satisfied only for $k>k_{0},$ see Eq.(\ref{eq:Boltzmann transport mean free path 2}).
The cutoff $k_{c}$, below which this condition fails, is now given
by $k_{c}\sim k_{0}\eta^{\frac{2}{5}}>k_{0}.$ Thus, $n(\vec{r},t)$
is given by the second term in (\ref{eq:Two integrals}), but with
the lower limit of integration being equal to $k_{c}$. 

Let us return to the question of validity of the expression (\ref{eq:Long time limit}).
It was argued that the transition from (\ref{eq:Density Fermi 2})
to (\ref{eq:E_0>2E_F one integral Final with mu}), i.e. the replacement
of $P_{p}(\vec{r},\vec{R},t)$ by $P_{p}(\vec{r},0,t)$ is justified
for sufficiently long time. However, whether a given time $t$ can
be considered {}``sufficiently long'' depends on the value of the
diffusion coefficient for the relevant particles. It is clear that
for {}``fast'' particles, which rapidly diffuse out from the vicinity
of the trap, (\ref{eq:E_0>2E_F one integral Final with mu}) will
become accurate at earlier times than for slow particles, which tend
to stay in the vicinity of the trap for much longer. Formally, the
replacement of $P_{p}(\vec{r},\vec{R},t)$ by $P_{p}(\vec{r},0,t)$
requires \begin{equation}
\frac{rR_{Max}}{2D(k)t}<1\label{eq:condition 1}\end{equation}
and\begin{equation}
\frac{R_{Max}^{2}}{4D(k)t}<1\:.\label{eq:condition 2}\end{equation}
For $r>R_{Max}$ it is sufficient to satisfy only (\ref{eq:condition 1}),
because (\ref{eq:condition 2}) will be satisfied automatically. Then,
for some fixed $r$ one can identify three different time limits.
For short times, $t<\frac{rR_{Max}}{2D(k_{F})t}$, (\ref{eq:condition 1})
breaks down. This, however, is of no consequence since at such small
times even the fastest particles have not yet arrived to point $r$
(more precisely, particle density there is exponentially small). For
intermediate time, $t\sim\frac{r^{2}}{4D(k_{F})}$, the fast particles
arrive to point $r$ and the above conditions are satisfied for these
particles (these conditions are not satisfied for slow particles but
this is irrelevant since, for these $r$ and $t$, the contribution
of slow particles to $n(r,t)$ is small). For longer times, $t>\frac{r^{2}}{4D(k_{F})},$
the fast particles ($k\sim k_{F}$) have already diffused away and
slower particles start to arrive at point $r$. The arrival time for
particles with a given value of $k$ (smaller than $k_{F}$) is of
order $\frac{r^{2}}{D(k)}$ so that the condition (\ref{eq:condition 1})
is satisfied for these particles. It follows, thus, that for $r\gg R_{Max}$
the above conditions are satisfied for the {}``relevant'' particles,
i.e. the ones which dominate the concentration at a given $r$ and
$t$. 

For $r<R_{Max}$ the more stringent condition is (\ref{eq:condition 2})
and in order for it to be satisfied for the smallest wave number $k=k_{c}$,
one needs $t>\frac{R_{max}^{2}}{4D(k_{c})}$, i.e. $\widetilde{t}>\frac{1}{\eta^{2}}$
(we assume here $\eta\ll1$). In order to obtain more accurate results
for $r<R_{Max}$ and for not too long times, one has to return to
Eq. (\ref{eq:Density Fermi 2}) and use the kernel $P_{p}(\vec{r},\vec{R},t)$,
rather than the long time approximation $P_{p}(\vec{r},0,t)$. It
turns out that for $r=0$ and for the case $k_{F}\ll k_{0}$ Eq.(\ref{eq:Density Fermi 2})
can be evaluated exactly for an arbitrary time $t$:\begin{equation}
\widetilde{n}(0,\widetilde{t})=4Ns\left[\eta^{2}\left(e^{\frac{\pi s}{\widetilde{t}}-\frac{1}{2\widetilde{t}\eta^{2}}}-1\right)+\frac{1}{2\widetilde{t}}e^{\frac{\pi s}{\widetilde{t}}}\left(E_{1}\left[\frac{\pi s}{\widetilde{t}}\right]-E_{1}\left[\frac{1}{2\widetilde{t}\eta^{2}}\right]\right)\right]\:.\label{eq: n(0,t) Final solution}\end{equation}
For $\widetilde{t}>\pi s$, (\ref{eq: n(0,t) Final solution}) can
be cast in the following form:\begin{equation}
\widetilde{n}(0,\widetilde{t})\approx\frac{2Ns}{\widetilde{t}}\left(\ln\left[\min\left[\frac{\widetilde{t}}{\pi s},\frac{1}{2\pi s\eta^{2}}\right]\right]+O\left(1\right)\right)\:.\label{eq: Compact with min}\end{equation}
It is instructive to compare the above results for $\widetilde{n}(0,\widetilde{t})$
with the solution for the constant diffusion coefficient $D=D_{<}(k_{F})$,
which, for $\widetilde{t}>\pi s$ , is approximately\begin{equation}
\widetilde{n}(0,\widetilde{t})\approx\frac{Ns}{\widetilde{t}}(1-2\pi s\eta^{2})^{2}.\label{eq:Dconst 2D r_0 large t cutoff}\end{equation}
In the case of the speckle disorder, the decay is slowed down by the
factor $2\ln\left[\min\left[\frac{\widetilde{t}}{\pi s},\frac{1}{2\pi s\eta^{2}}\right]\right]$,
reflecting slower diffusion of less energetic particles. As an illustration,
in Fig. 6 we compare these two cases for $s=12$ and $\eta=0.05$.
Note that for $\widetilde{t}>\frac{1}{\eta^{2}}$, (\ref{eq: n(0,t) Final solution})
reduces to Eq. (\ref{eq:t more 1/eta from approximation}). %
\begin{figure}[H]
\centering{}\includegraphics[scale=0.5]{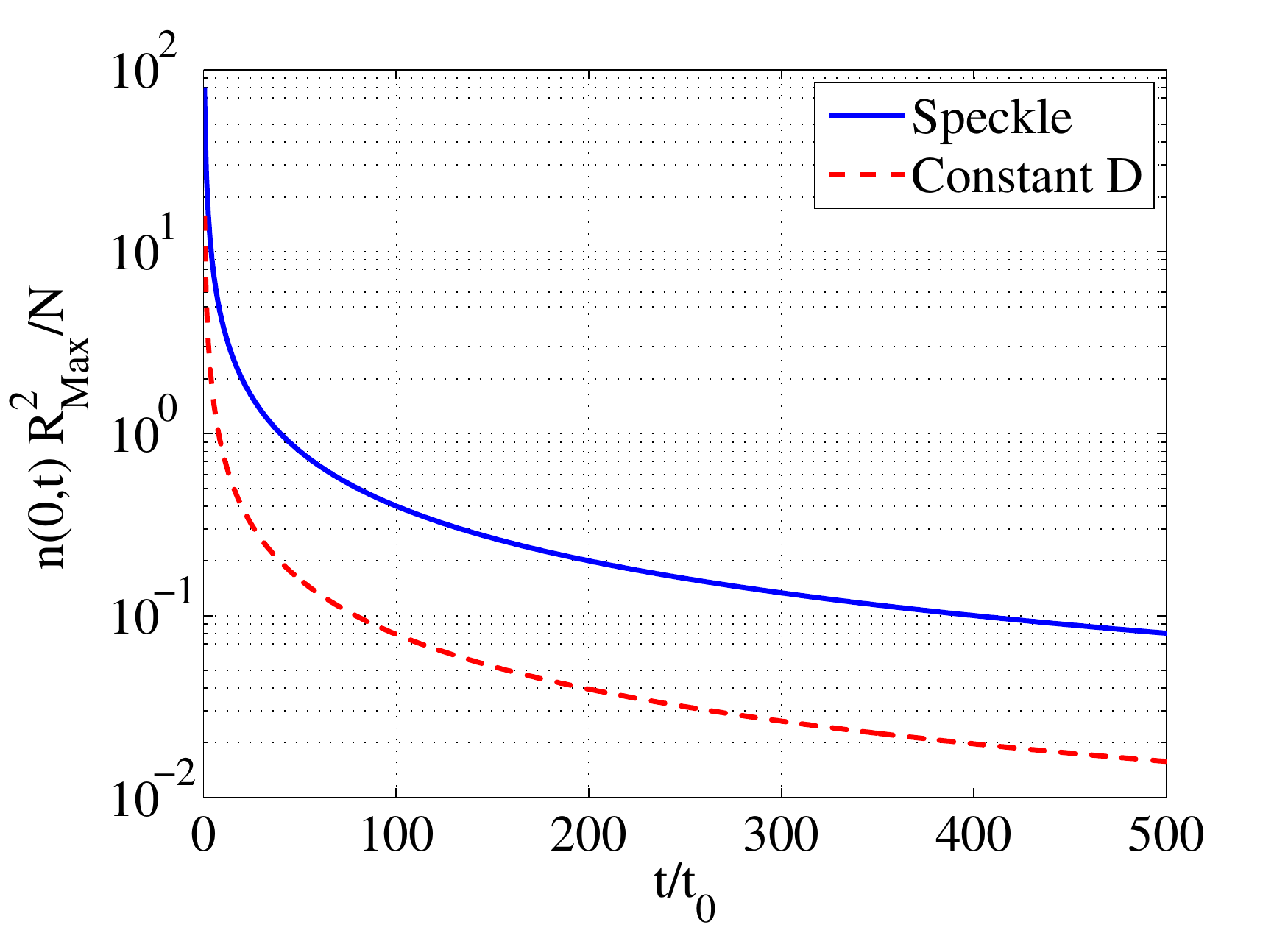}\caption{(Color online) Comparison between the Eq.(\ref{eq: Compact with min})
and the corresponding expression for a constant diffusion coefficient
Eq.(\ref{eq:Dconst 2D r_0 large t cutoff}), for $s=12$ and $\eta=0.05$. }

\end{figure}

For the case $k_{F}\gg k_{0}$ the expression for $\tilde{n}(0,\tilde{t})$
is more cumbersome and involves incomplete Gamma functions. The main
differences from the case $k_{F}\ll k_{0}$ occurs for times $s^{\frac{5}{2}}\ll\tilde{t}\ll1$.
For such times, the density decays as \begin{equation}
\tilde{n}(0,\tilde{t})\sim\tilde{t}^{-\frac{2}{5}}.\label{eq:E_less_r_0}\end{equation}
 For larger times, $\tilde{t}\gg1$, the behavior of the density $\tilde{n}(0,\tilde{t})$
will be generally similar to the case $k_{F}\ll k_{0}$, as discussed
above.

Finally, let us calculate the variance $\Delta r^{2}\left(t\right)=\int n(\vec{r},t)r^{2}d\vec{r}$
of the expanding density profile. Substituting $n(\vec{r},t)$ from
(\ref{eq:Density Fermi 2}), we obtain \begin{equation}
\Delta r^{2}\left(t\right)=\frac{d}{2+2d}R_{max}^{2}+2d\bar{D}t\,,\end{equation}
 where $\bar{D}$ denotes average of $D\left(p\right)$ over the momentum
distribution:\begin{equation}
\bar{D}=\frac{2\Gamma(d)}{\pi^{\frac{d}{2}}\Gamma(\frac{d}{2})}\intop_{\left|p\right|<p_{F}}\frac{d\vec{p}}{p_{F}^{2d}}\left(p_{F}^{2}-p^{2}\right)^{\frac{d}{2}}D\left(p\right)\,.\end{equation}
 Similar result was obtained in Ref.\cite{Miniatura-1} for the variance
of the BEC cloud expanding from the harmonic trap. In that case, however,
the momentum distribution is given by the inverted parabola in both
$d=2$ and $d=3$.

\section{Diffusion of a bec}

Previous sections were devoted to a cold Fermi gas. In this section
we briefly discuss diffusion of a BEC expanding through an optical
speckle. This problem has been addressed in a rather detailed and
experimentally relevant paper of Miniatura \textit{et al} \cite{Miniatura-1},
with an emphasis on the limiting stationary density distribution.
Here we concentrate on the earlier stages of the time evolution of
the expanding BEC cloud. Our treatment will be within the mean field
(Gross-Pitaevskii) approximation, when the BEC can be described by
a single macroscopic wave function $\Psi(\vec{r},t)$. The expansion
occurs in two stages, when the first stage is dominated by the nonlinearity
whereas the second stage describes a linear evolution in the presence
of disorder \cite{Shapiro-1,Miniatura-1,Skipetrov-1,Sanches}. Initially
the condensate is prepared in a harmonic trap (frequency $\omega$)
and its energy is dominated by interactions, i.e. by the nonlinear
term in the Gross-Pitaevskii equation. At time $t=0$ the trap is
switched off and the BEC undergoes a free (ballistic) expansion for
a time $t_{0}$ equal to few ($\frac{1}{\omega}$). By that time the
interaction energy, stored in the initial wave packet, is converted
into the kinetic energy of the condensate flow so that the interaction
can be neglected. At $t=t_{0}$ the random speckle potential is switched
on and the BEC evolves according to the linear Schrödinger equation,
with the static potential $V(\vec{r})$. It has to be solved with
the initial condition $\Psi(\vec{r},t_{0})=\Phi(\vec{r})$, where
$\Phi(\vec{r})$ is the condensate wave function at time $t_{0}$.
Its shape is given by an inverted parabola, with superimposed rapid
phase oscillations indicating large kinetic energy (see Eq.(23) of
Ref. \cite{Miniatura-1}). Measuring the time from the instant $t_{0}$,
the standard treatment leads to the following expression for the condensate
density, averaged over various realizations of $V(\vec{r})$ (compare
to (\ref{eq:diffusion expectation value averaged2})):\begin{equation}
n_{B}(\vec{r},t)=\int d\vec{R}\int d\vec{p}P_{p}(\vec{r},\vec{R},t)W_{B}(\vec{p},\vec{R})\:,\label{eq:Bosons expectation value}\end{equation}
where $W_{B}(\vec{p},\vec{R})$ is the Wigner function corresponding
to the wave function $\Phi(\vec{r})$. Let's compare (\ref{eq:Bosons expectation value})
with the corresponding expression (\ref{eq:Density Fermi 2}) for
fermions. Defining the {}``effective Wigner function'' of the Fermi
gas as\begin{equation}
W_{F}(\vec{p},\vec{R})=\frac{1}{(2\pi\hbar)^{d}}\Theta(E_{F}-\frac{p^{2}}{2m}-\frac{1}{2}m\omega^{2}R^{2})\end{equation}
we can write (\ref{eq:Density Fermi 2}) exactly in the form as (\ref{eq:Bosons expectation value}),
with $W_{B}$ replaced by $W_{F}$. The two functions have much in
common. Integration over $\vec{p}$ and over $\vec{R}$, respectively,
shows that the spatial distribution and the momentum distribution
for $W_{F}(\vec{p},\vec{R})$ are inverted parabolas (in $2d$), with
characteristic length $R_{Max}=\frac{\hbar p_{F}}{m\omega}$ and characteristic
momentum $p_{F}.$ But such inverted parabolas (with $k_{F}$ replaced
by the inverse healing length $1/\xi$ of the BEC prior to the release
from the trap) are well known to correspond to the condensate wave
function $\Phi(\vec{r})$ and, thus, to the Wigner function $W_{B}(\vec{p},\vec{R})$.
It is therefore clear that the dynamics of a BEC and of a Fermi cloud
must be quite similar. (This similarity has been used in \cite{Shapiro-1}
to propose a single parameter scaling for BEC dynamics). For instance,
in the long time limit discussed in Sec. $\Romannumber{4}$, when
$\vec{R}$ in the diffusion kernel $P_{p}(\vec{r},\vec{R},t)$ can
be set to zero, (\ref{eq:Bosons expectation value}) will involve
only the momentum distribution $\int d\vec{R}W_{B}(\vec{p},\vec{R})$
and, thus, the functional form of $n_{B}(\vec{r},t)$ will be identical
to that of the Fermi gas. Therefore all the results based on Eq.(\ref{eq:Long time limit})
- such as those given in (\ref{eq:White noise limit}) or (\ref{eq:Two integrals})
- hold also for a BEC (with the replacement $k_{F}\rightarrow1/\xi$).

One should keep in mind that, in spite of having much in common, the
functions $W_{B}(\vec{p},\vec{R})$ and $W_{F}(\vec{p},\vec{R})$
are not identical (indeed, two Wigner functions with the same spatial
and momentum distributions do not necessarily coincide!). Therefore,
for $r\lesssim R_{Max}$ (and for not too long times) the shape of
a BEC cloud is expected to differ significantly from that of a Fermi
gas. Eq.(\ref{eq:White noise limit}) is not applicable in this regime
and one should use the more elaborate Eq.(\ref{eq:Bosons expectation value})
which is the bosonic counterpart of Eq.(\ref{eq:Density Fermi 2})
for fermions.

\section{Conclusions}

We have considered diffusion of a Fermi gas in the presence of a random
optical speckle potential. The problem, although straightforward in
principle, is quite involved technically and it differs in several
respects from the standard diffusion problem encountered in condensed
matter physics \cite{Akkermans}. One difference is that a broad range
of particle momenta has to be considered, rather than a narrow interval
near the Fermi momentum (as is usually the case for the electronic
systems). Another difference is that the speckle potential has long
range correlations.

We have emphasized the importance of the parameter $\eta=\frac{V_{0}}{E_{0}}$,
where $V_{0}$ and $E_{0}$ are, respectively, the typical amplitude
and the {}``correlation energy'' of the potential \cite{Shklovskii}.
For $\eta\ll1$, particles with wave number $k<k_{0}$ do not feel
correlations in the potential and diffuse as in a white noise potential.
For $\eta\gg1$, on the other hand, correlations are important for
all particles, regardless of their momenta. In that case an accurate
estimate of the lower cutoff, $k_{c}$, below which classical Boltzmann
transport is impossible, becomes somewhat ambiguous. Our estimate
was based on the Ioffe-Regel criterion, $k_{c}l_{B}=1,$ and it leads
to $k_{c}\sim k_{0}\eta^{\frac{2}{5}}>k_{0}.$ This corresponds to
a critical energy $E_{c}\sim E_{0}\eta^{\frac{4}{5}}$ which is slightly
smaller than $V_{0}$. Since, however, $E_{c}$ is much above the
percolation threshold $E_{p}$ (in two dimensions $E_{p}=0$), there
exists a broad range of energies in which particles can propagate
by classical percolation (of course, in $2d$, and at sufficiently
large distance, quantum interference will eventually take over and
lead to localization). Such {}``percolating particles'' were not
accounted for in our treatment. This omission can be partially rectified
by treating $k_{c}$ as a phenomenological fitting parameter whose
value is determined from experiment.

Although the paper is devoted primarily to fermions, we have discussed
in the last section diffusion of a BEC. It turns out that, within
the Gross-Pitaevskii approximation, the shape of a diffusing BEC cloud
is remarkably similar to that of a Fermi gas.

All kinds of localization effects have been neglected in the present
paper, so that the weak disorder requirement, $k_{F}l_{B}\gg1,$ is
a necessary condition for the results to be valid. Finally, we have
focused on the $2d$ case. Similar calculations can be performed also
in $3d$, starting from Eq.(\ref{eq:Density Fermi 2}). Of course,
one has to use the $3d$ diffusion kernel and the appropriate expression
for the diffusion coefficient $D(k)$ in a $3d$ speckle potential.

\end{document}